\documentclass[letterpaper]{article}
\usepackage{aaai}
\usepackage{times}
\usepackage{helvet}
\usepackage{courier}

\usepackage{graphicx}
\usepackage{amsmath}

\usepackage{amsfonts}
\usepackage{booktabs}
\usepackage{url}

\usepackage{color}
\usepackage{xspace}

\usepackage[labelfont=bf,compatibility=false,font=small,labelsep=period]{caption
} 
\usepackage{subcaption}
\newcommand{\genderAcc}{82.3~\%\xspace}

\title{You Are What Apps You Use: \\  
Demographic Prediction Based on User's Apps}
\author{Eric Malmi\\
Verto Analytics and Aalto University\\
Espoo, Finland\\
eric.malmi@aalto.fi\\
\And Ingmar Weber\\
Qatar Computing Research Institute\\
Doha, Qatar\\
iweber@qf.org.qa\\
}
\begin{document}
\maketitle
\begin{abstract}
Understanding the demographics of 
app users is crucial, for example, for app developers, who wish to target their
advertisements more effectively.
Our work addresses this need by studying the predictability of user 
demographics based on the list of a user's apps which is readily available 
to many app developers. 
We extend previous work on the problem on three frontiers: 
(1)~We predict new demographics (age, race, and income) and analyze the most 
informative apps for four demographic attributes included in our  
analysis. The most predictable attribute is gender (\genderAcc accuracy), 
whereas the hardest to predict is income (60.3~\% accuracy).
(2)~We compare several dimensionality reduction methods for high-dimensional 
app data, finding out that an unsupervised method yields superior results 
compared to aggregating the apps at the app category level, but the best 
results are obtained simply by the raw list of apps.
(3)~We look into the effect of the training set size and the number of 
apps on the predictability and show that both of these factors have a large 
impact on the prediction accuracy. The predictability increases, or in other 
words, a user's privacy decreases, the more apps the user has used, but somewhat
surprisingly, after 100 apps, the prediction accuracy starts to decrease.
\end{abstract}

\section{Introduction}

%
In 2014, 60~\% of internet traffic was estimated to come from mobile 
devices\footnote{\url{
http://smallbiztrends.com/2014/07/online-traffic-report-mobile.html}}, of which 
51~\% was attributed to apps.\footnote{Another 2015 report by Morgan Stanley 
put this fraction closer to 33~\% though, with mobile web browsing dominating 
mobile traffic
\url{http://tinyurl.com/jstkty7}.
}
 As the importance of mobile apps continues to rise, some have even declared 
that ``apps are the new 
Web''\footnote{\url{http://tinyurl.com/3a3ru2o}}. Though claims of the Web's 
demise are probably exaggerated, the number of available mobile apps continues 
to increase and with it, one would expect, the importance of apps for the wider 
Web ecosystem.

At the same time, most academic studies looking at ``online users'' still 
concentrate on website visits with, by comparison, much fewer attention being 
given to mobile apps. In this paper, we study the predictability of six 
demographic attributes
based on the list of used apps.

The most apparent applications of demographic prediction methods are in 
marketing. For instance, an app developer might be interested in understanding 
which user segments are underrepresented when designing new ad campaigns for 
the app. 
On the other hand, computational social scientists, studying the behavior of 
people as observed through an app, like Twitter, need to understand how 
representative users of the app are as a sample of the whole population.

Studying the predictability of demographics also points out privacy 
implications of users allowing apps to access their list of installed apps. 
Many users undoubtedly do not carefully review the permissions that the apps 
they install require, and even less, understand the scope of the 
information that can be inferred from the data accessible by the apps.

The only previous studies on demographic prediction based on lists of apps, we 
are aware of, are \cite{seneviratne2014,seneviratne2015}. In the latter work, 
only gender prediction is studied and the dataset comprises of 218 users. We 
have obtained a dataset of 3\,760 users, which allows us to perform more 
fine-grained analyses, e.g., looking into the effect of app count on the 
predictability, and to obtain statistically more reliable results.

\section{Material}

Our dataset contains the demographic attributes and a list of apps for 3\,760 
Android users. While \cite{seneviratne2015} analyze the lists of 
\emph{installed} apps, we are studying the lists of apps \emph{used at least 
once} within a period of one month in 2015. Some very rarely used apps are 
probably missing from the latter list, but nevertheless, the lists can be 
expected to be highly correlated.

The average number of apps per user is 82.6 and the number of 
unique apps is 8\,840. Apps with less than ten users have been 
discarded to remove all personally identifiable information.
The dataset is from Verto Analytics
who have provided us a subsample of their media-measurement panel from the US. The 
panelists were recruited with the target of getting a representative 
sample of the US population. Each panelist has installed a meter app which 
tracks their app usage, and in return, the user is paid for providing the data.

\section{Methods}

When choosing a suitable prediction method, it is important to consider the 
following characteristics of the dataset: (1)~feature vectors 
(bags--of--apps) are binary and very sparse, (2)~the number of features, 
8\,840, is larger than the number of datapoints, 3\,760, and (3)~the 
dependent variables (user demographics) can be treated categorical.

Logistic regression is a natural choice for this type of a problem. We also 
tested support vector machines with different kernels and random forests, but 
both the 
results and the running times were inferior. While logistic regression can be 
adapted to multi-class problems, we instead binarize the demographic variables 
and balance the classes. This allows us to compare the predictability of 
different demographic variables.

\section{Results}

Next, we show results related to three different aspects of demographic 
prediction, namely, (1) the predictability of different demographics, (2) the 
effects of the training set size and the number of user's apps, and (3) the 
effect of various dimensionality reduction methods. 

\subsection{How Much is Revealed by Which Apps?}

Classification accuracies for six different demographic attributes are shown in 
Table~\ref{tab:dp}. They are computed based on a ten-fold 
cross-validation, and the most predictable attribute is \emph{gender}, whereas 
the \emph{household income} of a user is the most difficult to tell based on 
the list of user's apps. The results are surprisingly similar to the AUC scores 
reported by \cite{goel2012} who employed visited websites as the features 
instead of apps. In their work, the attributes marked with `*' had a slightly 
different binarization threshold compared to us.
The receiver operating characteristic for gender, given in 
Figure~\ref{fig:a}, shows that for half of the users, the gender can be 
predicted with a 97~\% accuracy.

\begin{table}[t]
\caption{Demographic prediction accuracy based on a user's apps. Classes have 
been binarized and balanced. AUC (Web) column shows the prediction performance 
based on visited websites from \cite{goel2012}.}\label{tab:dp}
\centering
\resizebox{\columnwidth}{!}{
\begin{tabular}{lllll}
\toprule
Attribute & Classes & Accuracy & AUC & AUC (Web)\\
\midrule
Gender & Male vs. Female & $\genderAcc$ & $0.901$ & $0.84$ \\
Age & 18--32 vs. 33--100 & $77.1~\%$ & $0.850$ & $0.85^*$ \\
Race & White vs. Non-white & $72.7~\%$ & $0.801$ & $0.83$ \\
Married & Married vs. Single & $72.5~\%$ & $0.792$ & NA \\
Children & 0 vs. $\geq 1$ children & $63.5~\%$ & $0.688$ & NA \\
Income & $\leq$ \$40K vs. $>$ \$40K & $60.3~\%$ & $0.645$ & $0.75^*$ \\
\bottomrule
\end{tabular}
}
\end{table}

By studying the coefficients of the logistic model, we can analyze the 
contribution of individual apps to the predictions. The coefficients with the 
largest absolute values are the most predictive ones. In 
Table~\ref{tab:predictors}, we have listed these apps for four
different demographics along with the coefficients, the shares and the numbers
of users who have used the app\footnote{Note that for a given app, the
number of users, $n$, might vary over demographics as some users have been
removed when balancing the classes.
}. Many of the results are not surprising; for instance, 
period tracking apps are good predictors for gender, whereas dating apps are 
more informative about the marital status. 
\begin{table*}[!htb]
\caption{The most predictive apps for different demographic attributes along with
the logistic regression coefficients (Coef), the fractions of app users with the
demographic attribute (Share), and the numbers of app users ($n$).
} \label{tab:predictors}
\centering
\resizebox{\textwidth}{!}{
\addtolength{\tabcolsep}{-1pt}
\begin{tabular}{llllllllllllllll}
\toprule


\multicolumn{4}{l}{{\bf Gender (Male)}} & \multicolumn{4}{l}{{\bf Age 
(33--100)}} & \multicolumn{4}{l}{{\bf Married (Married)}} & 
\multicolumn{4}{l}{{\bf Income ($\geq$ \$50K)}} \\ \midrule
Coef & Share & $n$ & App name & Coef & Share & $n$ & App name & Coef & Share & $n$ & App name & Coef & Share & $n$ & App name \\ \midrule
0.81 & 85~\% & 150 & ESPN & 0.53 & 80~\% & 42 & Great Clips Online Check-in & 0.55 & 67~\% & 200 & Zillow Real Estate \& Rentals & 0.58 & 75~\% & 141 & Fitbit \\
0.73 & 80~\% & 142 & Geek - Smarter Shopping & 0.48 & 53~\% & 1687 & Email & 0.44 & 67~\% & 622 & Walmart & 0.45 & 66~\% & 205 & LinkedIn \\
0.63 & 78~\% & 277 & Tinder & 0.46 & 58~\% & 318 & New Words With Friends & 0.44 & 60~\% & 823 & Pinterest & 0.41 & 65~\% & 41 & com.ws.dm \\
0.59 & 80~\% & 172 & Fallout Shelter & 0.44 & 80~\% & 65 & BINGO Blitz & 0.44 & 74~\% & 39 & Gospel Library & 0.37 & 52~\% & 141 & LG Android QuickMemo+ \\
0.56 & 86~\% & 106 & WatchESPN & 0.43 & 60~\% & 380 & iHeartRadio - Music \& Radio & 0.40 & 59~\% & 91 & USAA Mobile & 0.37 & 58~\% & 191 & Redbox \\
0.52 & 72~\% & 190 & Clash of Clans & 0.41 & 54~\% & 197 & Field Agent & 0.40 & 80~\% & 63 & ClassDojo & 0.36 & 72~\% & 22 & Like Parent \\
0.52 & 97~\% & 41 & Grindr - Gay chat, meet \& date & 0.40 & 55~\% & 690 & Lookout Security \& Antivirus & 0.38 & 60~\% & 123 & ESPN & 0.34 & 66~\% & 63 & Peel Smart Remote \\
0.49 & 84~\% & 96 & Yahoo Fantasy Football \& More & 0.40 & 92~\% & 41 & DoubleUCasino & 0.37 & 82~\% & 28 & Deer Hunter 2014 & 0.34 & 61~\% & 220 & Yelp \\
\midrule
\multicolumn{4}{l}{{\bf Gender (Female)}} & \multicolumn{4}{l}{{\bf Age 
(18--32)}} & \multicolumn{4}{l}{{\bf Married (Single)}} & 
\multicolumn{4}{l}{{\bf Income ($\leq$ \$40K)}} \\ \midrule
-1.03 & 76~\% & 736 & Pinterest & -1.17 & 78~\% & 1066 & Snapchat & -0.89 & 70~\% & 810 & Snapchat & -0.43 & 66~\% & 136 & Job Search \\
-0.73 & 84~\% & 182 & Etsy & -0.52 & 59~\% & 113 & Perk Word Search & -0.78 & 89~\% & 114 & POF Free Dating App & -0.43 & 63~\% & 97 & Security policy updates \\
-0.61 & 97~\% & 79 & Period Tracker & -0.49 & 64~\% & 88 & Summoners War & -0.73 & 85~\% & 219 & Tinder & -0.37 & 78~\% & 23 & Solitaire \\
-0.54 & 96~\% & 58 & Period Calendar / Tracker & -0.46 & 59~\% & 98 & Clash of Kings & -0.66 & 98~\% & 69 & OkCupid Dating & -0.35 & 67~\% & 79 & Prize Claw 2 \\
-0.50 & 76~\% & 346 & Cartwheel by Target & -0.45 & 86~\% & 90 & iFunny :) & -0.48 & 72~\% & 269 & Tumblr & -0.34 & 72~\% & 51 & ScreenPay- Get Paid to Unlock \\
-0.49 & 66~\% & 258 & Wish - Shopping Made Fun & -0.45 & 81~\% & 158 & GroupMe & -0.42 & 72~\% & 205 & SoundCloud - Music \& Audio & -0.33 & 78~\% & 56 & MeetMe \\
-0.49 & 74~\% & 325 & Checkout 51 - Grocery Coupons & -0.42 & 80~\% & 68 & GIPHY for Messenger & -0.41 & 65~\% & 331 & Uber & -0.33 & 62~\% & 77 & Foursquare \\
-0.45 & 74~\% & 178 & Photo Grid - Collage Maker & -0.42 & 80~\% & 183 & Vine & -0.41 & 89~\% & 69 & MeetMe & -0.32 & 56~\% & 73 & Microsoft Word \\

\bottomrule
\end{tabular}
\addtolength{\tabcolsep}{1pt}
}
\end{table*}

\subsection{How Much Does the Size Matter?}

\begin{figure*}[tb]
  \hspace*{\fill}
  \begin{subfigure}[t]{.33\textwidth}
    \centering
\includegraphics[width=\textwidth]{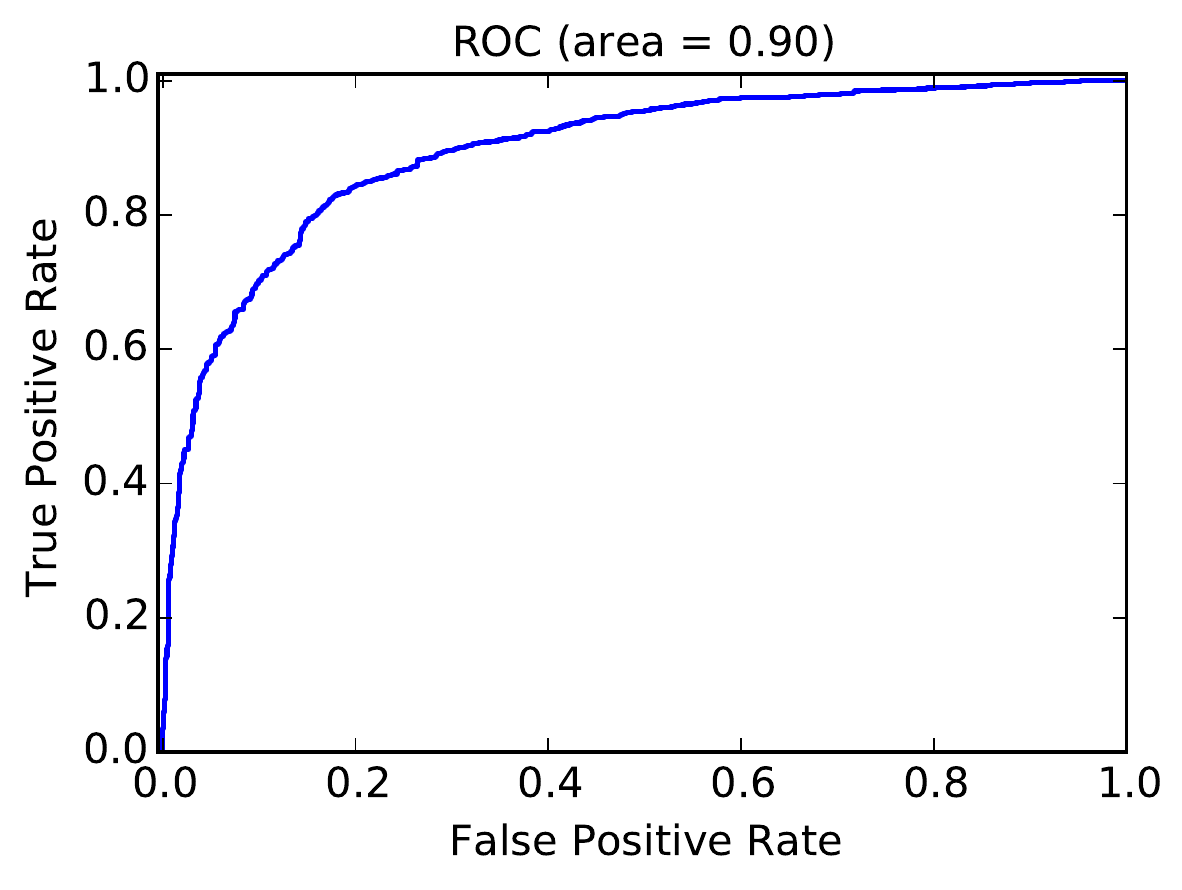}
    \caption{ROC curve for gender prediction. 'Male' is treated as the positive 
class.}
    \label{fig:a}
  \end{subfigure}%
  \hspace*{\fill}
  \begin{subfigure}[t]{.32\textwidth}
    \centering
    \includegraphics[width=\textwidth]{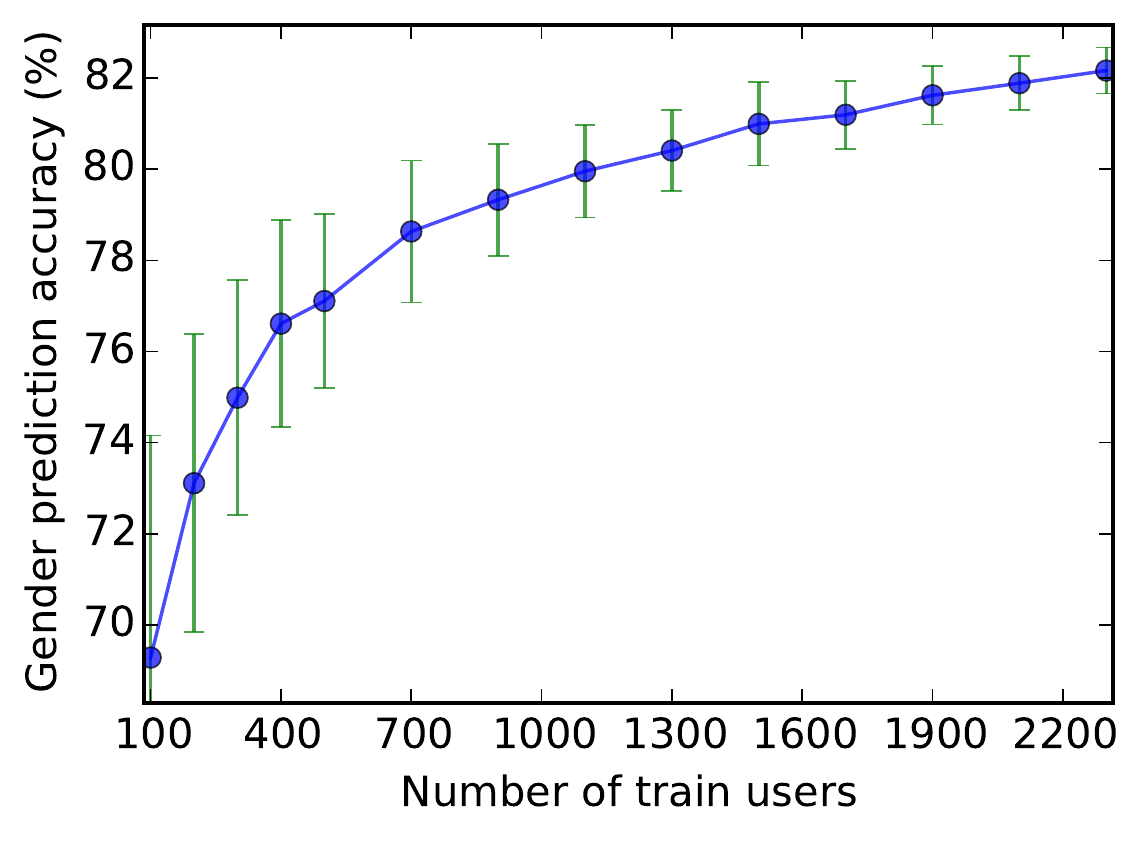}
    \caption{Effect of training set size on gender prediction.}
    \label{fig:b}
  \end{subfigure}%
  \hspace*{\fill}
  \begin{subfigure}[t]{.33\textwidth}
    \centering
    \includegraphics[width=\textwidth]{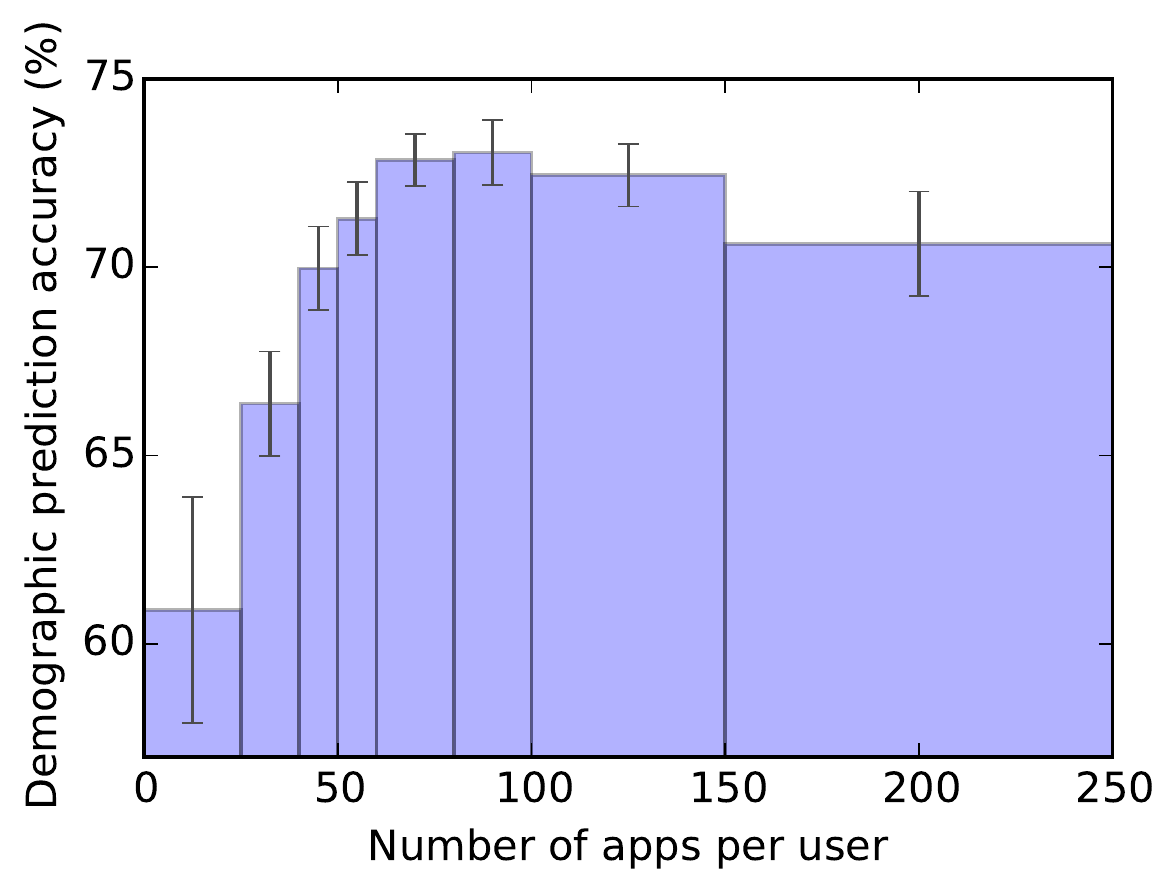}
    \caption{Effect of user's app count averaged over all demographics.}    
    \label{fig:c}
  \end{subfigure}%
  \hspace*{\fill}
  \caption{Demographic prediction results.} 
\end{figure*}

We also study the gender prediction accuracy as a function of the training 
set size in Figure~\ref{fig:b}. An absolute improvement of more than ten 
percents can be obtained by increasing the training set size from 100 users to 
2\,300 users. Error bars show the standard deviations of the accuracies over 
100 balanced random subsamples per given number of train users.


\cite{seneviratne2015} report an accuracy of 69.8~\% for a dataset of 174 
users, 50~\% of which are used for training (their original dataset is 218 
users but they undersample the majority class to balance the classes). To 
be able to benchmark against this result, we take 300 balanced random 
subsamples of 174 users and run 2-fold cross-validation for these samples. We 
obtain a comparable average accuracy of (68$\pm$5)\% even though we are not 
using content-based features derived from app descriptions nor numeric 
features as done in \cite{seneviratne2015}. This suggests that the bag--of--apps 
features alone can provide a competitive performance. Furthermore, these 
features can be extracted more easily, without having an access to an API for 
scraping the Google Play store.

A user who has installed a hundred apps probably reveals more of herself than a 
user with five apps. Thus it is relevant to ask, how quickly is privacy lost 
when the number of apps increases. In Figure~\ref{fig:c}, we tackle 
this question by binning the users according to the number of apps they have 
used and showing the prediction accuracy averaged over all demographic 
attributes of all users in a bin.
The standard errors are given by $\sqrt{p(1-p)/n}$.
The results show that the accuracy increases 
by about ten percents going from 20 apps to 100 apps but after that, somewhat
surprisingly, the accuracy starts to decrease. To test whether the decrease is
statistically significant, we perform an independent two-sample $t$~test with
the following null hypothesis: ``The overall demographic prediction accuracy is
not higher for the users with 50-150 apps compared to the users with more than
150 apps.'' We can reject the null hypothesis ($p=0.014$), which shows that
using a lot of apps at least once per month actually increases privacy.

\subsection{Dimensionality Reduction}

Due to the high dimensionality of feature vectors (8\,840 unique apps) we study 
three different dimensionality reduction approaches.

The first method, adopted by \cite{seneviratne2015}, considers only the 
apps installed by at least 10~\% of the users (125 apps in total). With this 
approach the gender prediction accuracy drops from \genderAcc to 73.6~\%, and  
even with the dataset of 174 users, the accuracy is decreased from 68~\% to 
65~\%. It is important to use the full list of apps since some of the apps 
might be reliable predictors even though they are rare (think, for example, of 
a rare \emph{period tracking} app).

The second method, also adopted by \cite{seneviratne2015}, aggregates the 
installed apps to category level based on Google Play categorization. In our 
dataset, there are apps from 48 categories. We take the number of apps in 
each category as the features, which yields an accuracy of 74.6~\%.

The third method employs the Truncated Singular Value Decomposition (TSVD). 
\cite{hu2007} also employ TSVD, but instead of using the SVD components 
directly as features for predicting the demographics of web users, they adopt a 
recommender system approach.
Setting the number of dimensions to 48, we obtain a gender prediction accuracy 
of 76.9~\%. This shows that rather than using the Google Play categories of the 
apps, it is better to use the same number of SVD components learned in an 
unsupervised manner. However, the performance is clearly worse compared to not 
using any dimensionality reduction, and even by increasing the number of 
SVD components, we were unable to exceed the performance of the logistic 
regression with all features, although with 500 components, the accuracy is 
already 81.8~\%.

In conclusion, none of the explored dimensionality reduction methods helped us 
to improve the gender prediction accuracy. We should also note that although 
TSVD can help to reduce the data dimensionality to about one-tenth of the 
original without losing much in accuracy, this still does not necessarily help 
with the space complexity of the method. The reason is that unlike the SVD 
components, the original bag--of--apps features are very sparse and the logistic 
regression implementation we use\footnote{\url{
http://scikit-learn.org/stable/modules/generated/sklearn.linear_model.LogisticRe
gression.html}} supports sparse matrices.

\section{Related Work}

Characterizing the demographics of Twitter users has been studied by
\cite{mislove2011} who infer geography, gender, and race of the users based 
on self-reported locations and the names of the users. They find large deviances 
from the demographic distribution of the overall population. \cite{pew2015} 
provide a more extensive demographic comparison of five social media platforms 
based on telephone interviews.
\cite{goel2012} look into the demographics and behavior of web users, whereas 
\cite{weber2011} study the same for search-engine users.

The demographic prediction based on user's apps has been previously studied by 
\cite{seneviratne2015} who predict the users' gender. In their previous work 
\cite{seneviratne2014}, they also predict language, country, relationship 
status, and whether the user is a parent, but instead of predicting these 
attributes directly, they first predict which apps are associated with the 
attributes and then check whether a user has apps corresponding to a given 
demographic attribute.
We extend these works by studying new demographics (age, race, 
and income), showing that increasing the training dataset size 
drastically improves the prediction accuracy, and comparing various 
dimensionality reduction methods for the app data.

Others have studied demographic prediction, e.g., based on website visits 
\cite{hu2007,goel2012}, social network features \cite{brea2014,al2012}, call 
patterns \cite{sarraute2014}, Twitter followers \cite{culotta2015} and 
profiles \cite{chen2015}, and location data \cite{riederer2015}. Related to 
demographic prediction, \cite{chittaranjan2013} investigate the predictability 
of personality traits based on apps and other smartphone usage features.
There also was an app for predicting personality based on installed
apps\footnote{\url{http://www.
idigitaltimes.com/what-do-your} \url{-apps-say
-about-you-new-app-iphone-here-tell} \url{-you
-410883}}.

\section{Conclusions and Discussion}

We studied the demographic prediction problem based on the list of used apps. 
Large differences in the predictability were observed between the six 
demographic attributes studied in this work, gender being the most predictable 
and income being the hardest to predict. The apps contributing the 
most to the predictions were identified for each attribute, revealing 
some expected patterns: dating apps are used, although not exclusively, by 
single people, and high-income people are more likely to use \emph{LinkedIn}, 
whereas lower-income people prefer an app called \emph{Job Search}.

We also studied various dimensionality reduction methods for high-dimensional 
app data (8\,840 unique apps), finding out that SVD yields 
superior results compared to aggregating the apps on app category level, but the 
best results are obtained simply by the raw list of apps.
Finally, we looked into the effect of the training set size and the number of 
apps on the predictability and showed that both of these factors can have an 
impact of over 10~\% on the prediction accuracy. Interestingly, the 
predictability increases the more apps the user has used, but after 100 apps, 
the prediction accuracy starts to decrease. The accuracy drop from users with 
50-150 apps to users with more than 150 apps was found to be statistically 
significant.

Several interesting questions are left for future work. First, we note that 
demographic attributes are most likely not independent, and therefore, 
predicting the attributes simultaneously, employing multi-label prediction 
techniques, could improve the performance.
Second, we plan to study the demographics of various popular apps to understand 
potential biases in their userbases compared to the whole population.
Third, it would be 
interesting to study the usage patterns of different demographic groups (as done 
previously in the context of web search \cite{weber2010}) to better understand 
the effects of demographic biases.

\section{Acknowledgments}
We would like to thank Verto Analytics for providing the dataset, Timo 
Smura for useful discussions, and Janaki Koirala for conducting some of the 
initial experiments.
\bibliographystyle{aaai}
\bibliography{sources}

\end{document}